\documentclass[conference]{IEEEtran}

\IEEEoverridecommandlockouts

\usepackage{cite}
\usepackage{graphicx}
\usepackage[cmex10]{amsmath}
\usepackage{algorithmic}
\usepackage{array}
\usepackage{mdwmath}
\usepackage{mdwtab}
\usepackage{stfloats}
\usepackage{url}

\begin{document}

\title{Parametric Channel Estimation by Exploiting \\ Hopping Pilots in Uplink OFDMA}

\author{\IEEEauthorblockN{Xiaochuan~Zhao, Tao~Peng and Wenbo~Wang}%
\IEEEauthorblockA{Wireless Signal Processing and Network Lab \\
Key Laboratory of Universal Wireless Communication, Ministry of Education \\
Beijing University of Posts and Telecommunications, Beijing, China \\
Email: zhaoxiaochuan@gmail.com}%
\thanks{This work is sponsored in
part by the National Natural Science Foundation of China under grant
No.60572120 and 60602058, and in part by the national high
technology researching and developing program of China (National 863
Program) under grant No.2006AA01Z257.}}

\maketitle

\begin{abstract}
%\boldmath
In this paper, a parametric channel estimation algorithm applicable
to uplinks of orthogonal frequency division multiple access (OFDMA)
systems whose subcarriers are pseudo-randomly allocated is proposed.
By exploiting pilot hopping, estimation of signal parameters via
rotational invariance technique (ESPRIT) is employed to estimate the
path delays of the sparse multipath fading channel. From the delay
information, a channel interpolator utilizing global pilots, which
can be irregular distributed, is derived to estimate the channel
state information on the desired tones. Moreover, a simple method of
estimating the time correlation of the channel taps is introduced
and integrated in the proposed algorithm. Simulation results
demonstrate that the proposed algorithm outperforms the local linear
channel interpolator within a wide range of SNR's and Doppler's.
\end{abstract}


\IEEEpeerreviewmaketitle

\section{Introduction}
\label{sec:intro}

Recently, the pseudo-random resource allocation scheme is adopted by
wireless standards such as IEEE 802.16e wireless MAN standard
\cite{WiMAX16e} to provide frequency diversity and co-channel
interference averaging in orthogonal frequency division multiple
access (OFDMA) systems. However, the price paid for the flexibility
of such complex allocation scheme is the increased difficulty in
estimating the channel impulse response (CIR) or channel frequency
response (CFR) of the uplink. Since the pilot tones allocated to a
certain user are pseudo-randomly interleaved with the others,
traditional channel interpolators
\cite{Edfo98}\cite{Cole02}\cite{Dong07} using global pilots are
impracticable, while the local linear interpolators suffer from
significant performance degradation at high SNR regime where the
estimation error floor shows itself.

In order to overcome the irregular pilot pattern, \cite{Ragh07} has
proposed a parametric channel estimation \cite{Yang01} scheme which
can greatly reduce the channel estimation error for sparse multipath
channels. \cite{Ragh07} has applied estimation of signal parameters
via rotational invariance technique (ESPRIT) \cite{Roy89} which
exploits the shift-invariance property of uplink tile structure of
IEEE 802.16e to estimate and track multipath delay spread of the
received signal. With the multipath delays, the channel frequency
correlation information can be derived to enable the sinc-function
interpolator to estimate the CFR on the desired tones.

The downside of \cite{Ragh07} is its dependence on the special tile
structure which introduces the shift-invariance property --- the
base of ESPRIT, however this symmetric structure may be corrupted.
For example, when uplink virtual multiple-input multiple-output
(MIMO) is active, pilots in each tile are divided into two exclusive
subsets and allocated to two users, respectively. For any one, the
pilots in the subset lose the shift-invariance property. As a
result, \cite{Ragh07} fails under uplink virtual MIMO.

In this paper, the strict restriction of the special symmetry
structure of pilot pattern is relaxed, in stead, pilot hopping is
exploited to enable the parametric channel estimation under
arbitrarily irregular pilot distribution. By combining two vectors
of pilots of two contiguous OFDMA symbols, the shift-invariance
property is recovered. The sample auto-correlation matrix of the
combined pilot vectors is formed, and from which, time correlation
of the multipath fading channel is estimated to compensate the time
variance induced by Doppler. Minimum description length (MDL)
\cite{Wax85} is adopted to identify the number of significant paths.
With the delay information, a global channel interpolator is
introduced to estimate the CFR on the desired tones.

The rest of this paper is organized as follows. Section II
introduces the channel model and the OFDMA system model. The
proposed channel estimation algorithm is presented in Section III.
Simulation results and analysis are provided in Section IV. Finally,
Section V concludes the paper.

\emph{Notation}: Boldface letters denote matrices or column vectors.
$tr(\cdot)$, $(\cdot)^*$, $(\cdot)^H$, $(\cdot)^{\dag}$, and
$||\cdot||_F$ denote trace, conjugate, conjugate transposition,
Moore-Penrose pseudo inverse, and Frobenius norm, respectively.
$E(\cdot)$ represents the expectation of a stochastic process.
$[{\bf{\cdot}}]_{i,j}$ denotes the $(i$,$j)$-th element of a matrix.
$span(\bf{A})$ denotes the column space spanned by $\bf{A}$.

\section{System Model}
\label{sec:model}

Consider the uplink of an OFDMA system with a bandwidth of $BW =
1/T$ Hz ($T$ is the sampling period). We use $N$ to denote the total
number of tones and $N_{used}$ the total number of useful tones,
including data and pilot tones. A cyclic prefix (CP) of length
$L_{cp}$ is inserted before each OFDMA symbol to eliminate
inter-block interference. Thus the whole symbol duration is $T_s =
(N+L_{cp})T$.

Now we consider a certain user is scheduled to transmit over $M$ ($M
\leq N_{used}$) tones per symbol. The indexes of these tones are
collected in a set, denoted as ${{\mathcal{I}}}_0(n)$, which can be
changed along the time. Regardless of CP, the transmitted signal of
this user can be expressed in matrix form as
${{\bf{x}}(n)}={\bf{F}}^H{\bf{s}}(n)$, where ${\bf{s}}(n) \in
\mathcal {C}^{N \times 1}$ has non-zero elements at
${\mathcal{I}}_0(n)$, and ${\bf{F}} \in \mathcal {C}^{N \times N}$
is the balanced Fourier transform matrix with the $(m,n)$-th entry
$\frac{1}{\sqrt{N}}e^{-j2{\pi}mn/N}$.

The discrete complex baseband representation of a mobile wireless
CIR of length $L$ can be described by \cite{Stee92}
\begin{equation}
{h(n,\tau)=\sum\limits_{l=0}^{L-1}{\gamma_l(n)\delta\left({\tau-\tau_l}\right)}}
\end{equation}
where $\tau_l \in {\mathcal{R}}$ is the non-sample-spaced delay of
the $l$-th path normalized by the sampling period $T$, and
$\gamma_l(n)$ is the corresponding complex amplitude. Due to the
motion of mobile stations (MS's), $\gamma_l(n)$'s are wide-sense
stationary (WSS) narrowband complex Gaussian processes, which are
uncorrelated among different paths based on the assumption of
uncorrelated scattering (US). Furthermore we assume delays of
multipath are static during the process of estimation, as they are
varying very slowly compared with the complex amplitude of paths
\cite{Tse05}. We collect $\tau_l$'s into a set, denoted as
${\mathcal{I}}_{\tau}$.

In addition, We assume that $\gamma_l(n)$'s have the same normalized
correlation function $r_t(m)$ for all $l$'s, hence
\begin{equation}
{r_{\gamma_l}(m)=E\left[{\gamma_l(n+m)\gamma_l^*(n)}\right]=\sigma_l^2r_t(m)}
\end{equation}
where $\sigma _l^2$ is the power of the $l$-th path.

Assuming a sufficient CP, i.e., $L_{cp} \geq \tau_{L-1}$, the
received signal of the base station (BS) for the given user, denoted
as ${\bf{y}}(n) \in \mathcal {C}^{N \times 1}$, can be described as
\begin{equation}
{{\bf{y}}(n) = {\bf{X}}(n){\bf{H}}(n) + {\bf{v}}(n)}
\end{equation}
where ${\bf{X}}(n) \in \mathcal{C}^{N \times N}$ is a diagonal
matrix with non-zero elements drawn from ${\bf{x}}(n)$, ${\bf{H}}(n)
\in \mathcal{C}^{N \times 1}$ is CFR experienced by the $n$-th OFDMA
symbol, and is expressed as
\begin{equation}
{{\bf{H}}(n) = {{\bf{F}}_{\tau}{\bf{h}}}(n)}
\end{equation}
where ${\bf{h}}(n)=\left[h(n,\tau_0),...,h(n,\tau_{L-1})\right]^T$,
and ${\bf{F}}_{\tau} \in \mathcal{C}^{N \times L}$ is the
non-balanced Fourier transform matrix with the $(k,l)$-th entry
$e^{-j2{\pi}k{\tau}_l/N}$, and ${\bf{v}}(n) \in \mathcal {C}^{N
\times 1}$ is the zero-mean complex Gaussian noise vector with
variance $\sigma_n^2$, i.e., ${\bf{v}}(n) \sim \mathcal
{CN}(0,\sigma_n^2 {\bf{I}}_N)$.

\section{Channel Estimation}
\label{sec:chanEst}

Denote the number of pilots for the given user as $P \leq M$, the
indexes of pilots are collected into a set denoted as
${\mathcal{I}}_p(n) \subset {\mathcal{I}}_0(n)$. At the BS, pilots
are firstly extracted after Fourier transformation to perform
least-squared (LS) channel estimation in frequency domain, i.e.,
\begin{equation}
{{\bf{\hat H}}_p(n) = {\bf{X}}_p^{-1}(n){\bf{y}}_p(n) =
{\bf{H}}_p(n) + {\bf{w}}_p(n)}
\end{equation}
where, ${\bf{X}}_p(n) \in \mathcal{C}^{P\times{P}}$ is the pre-known
pilot diagonal matrix, ${\bf{\hat H}}_p(n) \in \mathcal{C}^{P \times
1}$ is the estimated CFR, and ${\bf{w}}_p(n) \in \mathcal{C}^{P
\times 1}$ is the noise vector expressed as
\begin{equation}
{{\bf{w}}_p(n) = {\bf{X}}_p^{-1}(n){\bf{v}}_p(n)}
\end{equation}
and ${\bf{w}}_p(n) \sim \mathcal{CN}(0, \sigma_n^2{\bf{I}}_P)$ when
${\bf{X}}_p^H {\bf{X}}_p = {\bf{I}}_P$ for PSK modulated pilots.

In the following section, we will introduce a simple pilot hopping
scheme based on which the parametric channel estimation can be
carried out.

\subsection{Pilot Hopping Pattern}
\label{ssec:ph}

Pilot hopping is made in two phases: the inner hopping between two
contiguous OFDMA symbols and the outer hopping superposed over the
inner one.

First, the inner hopping is carried out, which shifts all pilots of
odd OFDMA symbols by a pre-defined offset denoted as $\nu$.
Therefore, after that, ${\mathcal{I}}_{p,od}-{\mathcal{I}}_{p,ev} =
\nu$, where ${\mathcal{I}}_{p,od}$ and ${\mathcal{I}}_{p,ev}$ are
indexes of pilots of odd and even OFDMA symbols, respectively.
Second, the outer hopping is performed by shifting the pilots of
both even and odd symbols with the same offset.

For the proposed channel estimation algorithm, the inner hopping is
indispensable, while the outer hopping is optional --- although it
can accelerate the estimation of the auto-correlation matrix of CFR
by simulating more channel fading at the receiver through including
phase rotation \cite{Ragh05}. For simplicity, the outer hopping is
not applied in this paper. Later on, the shift-invariance structure
is exploited from the inner pilot hopping scheme.

\subsection{Estimation of Multipath Delays}
\label{ssec:sip}

Stack two contiguous estimated CFR vectors into one with the even on
the odd, i.e.,
\begin{equation}
{\bf{\hat H}}_{con,p}(n) = \left[ \begin{array}{l}
 {\bf{\hat H}}_p(2n) \\
 {\bf{\hat H}}_p(2n+1) \\
 \end{array} \right] = {\bf{H}}_{con,p}(n) + {\bf{w}}_{con,p}(n)
\end{equation}
where
${\bf{H}}_{con,p}(n),{\bf{\hat{H}}}_{con,p}(n)\in\mathcal{C}^{2P\times1}$
are the concatenated CFR vector and its estimation, respectively,
and ${\bf{w}}_{con,p}(n)\in\mathcal{C}^{2P\times1}$ is the
corresponding noise vector, and
${\bf{w}}_{con,p}\sim\mathcal{CN}(0,\sigma_n^2{\bf{I}}_{2P})$.

From (4), we have
\begin{equation}
{{\bf{H}}_p(2n)={\bf{F}}_{p,ev}{\bf{h}}(2n)}
\end{equation}
\begin{equation}
{{\bf{H}}_p(2n+1)={\bf{F}}_{p,od}{\bf{h}}(2n+1)}
\end{equation}
where ${\bf{F}}_{p,ev},{\bf{F}}_{p,od}\in\mathcal{C}^{P\times L}$
are submatrices drawn from ${\bf{F}}_{\tau}$ with rows indexes
belonged to ${\mathcal{I}}_{p,ev}$ and ${\mathcal{I}}_{p,od}$,
respectively. $P\geq{L}$ is required so that ${\bf{F}}_{p,ev}$ and
${\bf{F}}_{p,od}$ are of full column rank. As
${\mathcal{I}}_{p,od}-{\mathcal{I}}_{p,ev}=\nu$, it is
straightforward that
\begin{equation}
{{\bf{F}}_{p,od}={\bf{F}}_{p,ev}{\bf{\Phi}}}
\end{equation}
where ${\bf{\Phi}}\in{\mathcal{C}}^{L\times{L}}$ is a diagonal
phase-twisted matrix with the $l$-th diagonal element
$[{\bf{\Phi}}]_{l,l}={e^{-j2{\pi}{\nu}{\tau}_l/N}}$. Apparently,
${\bf{\Phi}}$ contains the multipath delay information $\tau_l$'s as
expected.

According to (8)-(10), (7) can be rewritten into
\begin{eqnarray}
\lefteqn{ {\bf{\hat H}}_{con,p} \left( n \right)} \nonumber \\
{} & = & \left[ {\begin{array}{cc}
   {{\bf{F}}_{p,ev} } & {\bf{0}}  \\
   {\bf{0}} & {{\bf{F}}_{p,ev} {\bf{\Phi }}}  \\
\end{array}} \right] %
\left[ \begin{array}{c}
 {\bf{h}}\left( {2n} \right) \\
 {\bf{h}}\left( {2n + 1} \right) \\
 \end{array} \right] + {\bf{w}}_{con,p} \left( n \right) \nonumber\\
\end{eqnarray}

\begin{figure*}[!b]
\label{eqn:eqn12}
\normalsize \vspace{4pt} \hrule %
{\setlength\arraycolsep{2pt}
\begin{eqnarray}
{\bf{\hat R}}_{con,p} &  =  & \left[ \begin{array}{cc}
   {{\bf{F}}_{p,ev} } & {\bf{0}}  \\
   {\bf{0}} & {{\bf{F}}_{p,ev} {\bf{\Phi }}}
   \end{array}
\right]  %
 \left[ {\begin{array}{cc}
   {E\left( {{\bf{h}}(2n){\bf{h}}^H(2n)} \right)} & {E\left( {{\bf{h}}(2n){\bf{h}}^H(2n+1)} \right)}  \\
   {E\left( {{\bf{h}}(2n+1){\bf{h}}^H(2n)} \right)} & {E\left( {{\bf{h}}(2n+1){\bf{h}}^H(2n+1)} \right)} \\
\end{array}} \right]   %
 \left[ {\begin{array}{cc}
   {{\bf{F}}_{p,ev}^H} & {\bf{0}}  \\
   {\bf{0}} & {{\bf{\Phi}}^H {\bf{F}}_{p,ev}^H }  \\
\end{array}} \right] + \sigma _n^2 {\bf{I}}_{2P} \nonumber \\
{} & = & \left[ \begin{array}{cc}
   {{\bf{F}}_{p,ev}} & {\bf{0}}  \\
   {\bf{0}} & {{\bf{F}}_{p,ev} {\bf{\Phi}}}
   \end{array}
\right]  %
 \left[ {\begin{array}{cc}
   {{\bf{R}}_h(0)} & {{\bf{R}}_h(-1)}  \\
   {{\bf{R}}_h(1)} & {{\bf{R}}_h(0)}  \\
\end{array}} \right]   %
 \left[ {\begin{array}{cc}
   {{\bf{F}}_{p,ev}^H } & {\bf{0}}  \\
   {\bf{0}} & {{\bf{\Phi }}^H {\bf{F}}_{p,ev}^H }  \\
\end{array}} \right] + \sigma _n^2 {\bf{I}}_{2P} \quad
\end{eqnarray}
}
\end{figure*}

Consequently, the auto-correlation matrix of the concatenated
estimation of CFR is given in (12) shown at the bottom of the next
page, where ${\bf R}_h(m)\in\mathcal{C}^{L\times{L}}$ denotes the
auto-correlation matrix of the CIR vector with the lag $m$. From the
WSSUS assumption and (2), the $(i,j)$-th element
$\left[{\bf{R}}_h(m)\right]_{i,j}=\sigma_i^2 r_t(m)\delta(i-j)$,
where $\delta(\cdot)$ is the Kronecker function. Therefore,
${\bf{R}}_h(m)$ can be written as
\begin{equation}
{{\bf{R}}_h(m)=r_t(m){\bf{R}}_h(0)}
\end{equation}
where ${\bf{R}}_h(0)$ is a diagonal matrix, and the $l$-th diagonal
element $\left[{\bf{R}}_h(0)\right]_{l,l}=\sigma_l^2$.

Discarding the noise component temporally, the useful component on
the right-hand side of (12) is rewritten into
\begin{eqnarray}
\lefteqn{{\bf{R}}_{con,p}} \nonumber \\
  {} &=& \left[ \begin{array}{cc}
  {{\bf{F}}_{p,ev}{\bf{R}}_{h}(0){\bf{F}}_{p,ev}^H} & {\eta}{{\bf{F}}_{p,ev}{\bf{R}}_{h}(0){\bf{\Phi}}^H{\bf{F}}_{p,ev}^H} \\
  {{\eta}{\bf{F}}_{p,ev}{\bf{\Phi}}{\bf{R}}_{h}(0){\bf{F}}_{p,ev}^H} & {{\bf{F}}_{p,ev}{\bf{\Phi}}{\bf{R}}_{h}(0){\bf{\Phi}}^H{\bf{F}}_{p,ev}^H} \\
\end{array} \right] \nonumber \\
  {} &{\buildrel\Delta\over{=}}& \left[ \begin{array}{cc}
  {{\bf{A}}_{11}} & {{\bf{A}}_{12}} \\
  {{\bf{A}}_{21}} & {{\bf{A}}_{22}} \\
\end{array} \right]
\end{eqnarray}
where $\eta=r_t(1)=r_t(-1)$, since $r_t(m)=r_t(-m)$ for the WSS
Gaussian process $h(n,\tau)$. If $\eta$ is pre-known --- for
example, when the maximum Doppler, denoted as $f_d$, can be
measured, according to the Jakes' model, $r_t(1)=J_0(2{\pi}f_dT_s)$,
where $J_0(\cdot)$ is the zeroth order Bessel function of the first
kind --- we can eliminate the effect of Doppler from
${\bf{R}}_{con,p}$ by
\begin{equation}
  {\bf{\widetilde R}}_{con,p} = \left[ \begin{array}{cc}
  {{\bf{A}}_{11}} & {{\bf{A}}_{12}/{\eta}} \\
  {{\bf{A}}_{21}/{\eta}} & {{\bf{A}}_{22}} \\
\end{array} \right]
\end{equation}
However, when $\eta$ can not be obtained in advance, we can
approximately evaluate $\eta$ by
\begin{equation}
\hat\eta  =
\sqrt{\frac{{\left\|{{\bf{A}}_{12}}\right\|_F^2+\left\|{{\bf{A}}_{21}}\right\|_F^2}}
{{\left\|{{\bf{A}}_{11}}\right\|_F^2+\left\|{{\bf{A}}_{22}}\right\|_F^2}}}
\end{equation}
Noting when the pilots are equispaced and $P$ is a factor of $N$,
which is an optimal case \cite{Ohno04}, (16) is accurate. In fact,
at this circumstance, ${\bf{F}}_{p,ev(od)}^H{\bf{F}}_{p,ev(od)} =
P{\bf{I}}_L$, therefore
\begin{eqnarray}
\left\|{{\bf{A}}_{11}}\right\|_F^2 =
\left\|{{\bf{A}}_{22}}\right\|_F^2 &=&
P^2\sum\limits_{l=0}^{L-1}{\sigma}^4 \\
\left\|{{\bf{A}}_{12}}\right\|_F^2 =
\left\|{{\bf{A}}_{21}}\right\|_F^2 &=&
{\eta}^2P^2\sum\limits_{l=0}^{L-1}{\sigma}^4
\end{eqnarray}
From (17)(18), (16) is an accurate estimator of $\eta$.

Denoting
${\overline{\bf{F}}}_p=\left[{\bf{F}}_{p,ev}^T,\left({{\bf{F}}_{p,ev}{\bf{\Phi}}}\right)^T\right]^T$,
(15) is rewritten into
\begin{equation}
{{\bf{\widetilde{R}}}_{con,p}={\overline{\bf{F}}}{\bf{R}}_h(0){\overline{\bf{F}}}^H}
\end{equation}
and with the noise component of (12), (19) is
\begin{equation}
{{\bf{\hat{\widetilde{R}}}}_{con,p}{\buildrel\Delta\over{=}}{\bf{\widetilde{R}}}_{con,p}+{\sigma_n^2}{\bf{I}}_{2P}={\overline{\bf{F}}}{\bf{R}}_h(0){\overline{\bf{F}}}^H}+{\sigma_n^2}{\bf{I}}_{2P}
\end{equation}

Now (20) is of the standard form to apply ESPRIT algorithm, which we
will briefly discuss in the following. Firstly, the eigenvalue
decomposition (EVD) is applied to
${\bf{\hat{\widetilde{R}}}}_{con,p}$. Then $L$ dominant eigenvalues
are distinguished and the corresponding eigenvectors are collected
in a matrix denoted as $\bf{U}$. As
$span({\bf{U}})=span({\overline{\bf{F}}}_p)$, there exists a
nonsingular matrix $\bf{T}$ such that
${\bf{U}}={\overline{\bf{F}}}_p{\bf{T}}$. Vertically splitting
$\bf{U}$ into two parts, i.e.,
${\bf{U}}=\left[{\bf{U}}_{up}^T,{\bf{U}}_{dw}^T\right]^T$, it
follows that ${\bf{U}}_{up}={\bf{F}}_{p,ev}{\bf{T}}$ and
${\bf{U}}_{dw}={\bf{F}}_{p,ev}{\bf{\Phi}}{\bf{T}}$. Accordingly we
have
\begin{equation}
{{\bf{U}}_{dw}={\bf{U}}_{up}{\bf{T}}^{-1}{\bf{\Phi}}{\bf{T}}={\bf{U}}_{up}{\bf{Q}}}
\end{equation}
where ${\bf{Q}}={\bf{T}}^{-1}{\bf{\Phi}}{\bf{T}}$ is similar with
$\bf{\Phi}$, in other words, they have the common eigenvalues. We
can solve ${\bf{Q}}$ from (21) by the linear least-squares (LS)
criterion or the non-linear total least-squares (TLS) criterion for
a better solution. When ${\bf{Q}}$ is obtained, its eigenvalues,
denoted as ${\lambda}_l$, are calculated through EVD. Then the tap
delays are estimated as
\begin{equation}
{{\hat{\tau}}_l=\frac{arg(\lambda_l^*)N}{2{\pi}{\nu}},l=0,...,L-1}
\end{equation}
where $arg(\lambda)$ denotes the phase angle of $\lambda$ in the
range $[0,2\pi)$. Finally, as the phase angle wraps around with a
period of $2\pi$, the multipath delay is uniquely identified only
when ${\tau}<N/{\nu}$, where $\tau$ is normalized by the sampling
period $T$.

\subsection{Estimation of the Number of Significant Paths}
\label{ssec:estNumPath}

In practical applications, the auto-correlation matrix
${\bf{\hat{R}}}_{con,p}$ in (12) is unattainable. With a finite
number of received symbols, the sample auto-correlation matrix is
obtained as
\begin{equation}
{{\bf{\hat{R}}}_{con,p}^{\prime}=\frac{2}{N_t}\sum\limits_{n=1}^{N_t/2}{\bf{\hat{H}}}_{con,p}(n){\bf{\hat{H}}}_{con,p}^H(n)}
\end{equation}
where $N_t$ is the number of sample OFDMA symbols.
${\bf{\hat{R}}}_{con,p}^{\prime}$ is an asymptotic unbiased
consistent estimator of ${\bf{\hat{R}}}_{con,p}$.

Since ${\bf{\hat{R}}}_{con,p}$ is affected by the Doppler, it is not
used in the estimation of the number of significant paths. Instead,
the auto-correlation matrix of ${\bf{\hat{H}}}_p$ in (5) is adopted.
With (14), it can be easily obtained through
${\bf{\hat{R}}}_{con,p}$ as
\begin{equation}
{{\bf{\hat{R}}}_p=({\bf{A}}_{11}+{\bf{A}}_{22}})/2
\end{equation}
where we note ${\bf{\Phi}}{\bf{R}}_h(0){\bf{\Phi}}^H={\bf{R}}_h(0)$.

When ${\bf{\hat{R}}}_{con,p}^{\prime}$ is available, the sample
auto-correlation matrix ${\bf{\hat{R}}}_p^{\prime}$ is obtained
through (24) accordingly, with which, then, MDL is performed to
estimate the number of significant paths, denoted as $\hat{L}$.

\subsection{Channel Interpolation}
\label{ssec:chanInpl}

After estimating the multipath delays, further modifications are
made to enhance the performance. Due to many potential factors, such
as the inaccurate estimation of $\eta$, finite observations of CFR
and additive noise, ESPRIT is impaired: the estimated multipath
delays fluctuate around their true values within a certain range,
which results in an incomplete path subspace spanned. In addition,
since the multipath delays are non-sample-spaced, all paths leak
their power to the samples nearby. And the closer a sample gets to a
path, the stronger it is influenced by the path.

Considering the fluctuating estimated delays as well as the power
leakage, we prefer to broaden the "observation window" around each
estimated path delay to capture most of its power, i.e., for each
estimated path ${\hat{\tau}}_l$, a capturing window, denoted as
${\mathcal{W}}_{{\hat{\tau}}_l}=\{{\lfloor{\hat{\tau}}_l\rfloor-{\beta}},...,{\lceil{\hat{\tau}}_l\rceil+{\beta}}\}$,
is included, where ${\beta}$ is a predefined parameter such that
$2{\beta}{\hat{L}}\leq{P}$. Through the simulation, we find
$\beta\leq{5}$ is sufficient for a wide range of SNR's and
Doppler's. Consequently, the expanded set of estimated multipath
delays ${\mathcal{I}}_{\hat{\tau}}$ is
\begin{equation}
{{\mathcal{I}}_{\hat{\tau}}=\left(\bigcup\limits_{l=0}^{{\hat{L}}-1}{{\mathcal{W}}_{{\hat{\tau}}_l}}\right)\bigcap{{\mathcal{I}}_{cp}}}
\end{equation}
where ${{\mathcal{I}}_{cp}}=\{0,...,L_{cp}-1\}$ denotes the range of
CP, hence limits the estimated multipath delays within it.

The parameter $\beta$ requires carefully designing, since there
exists a tradeoff: the larger $\beta$ is, the more energy of paths
is captured, meanwhile, the more noise is also introduced, and vice
versa.

The channel interpolators, denoted as ${\bf{G}}_{ev}$ and
${\bf{G}}_{od}$ for even and odd symbols, respectively, are given as
\begin{equation}
{{\bf{G}}_{ev(od)}={{\bf{F}}_{d,ev(od)}}{{\bf{F}}_{p,ev(od)}^{\dag}}}
\end{equation}
where ${{\bf{F}}_{d,ev(od)}}$ is a matrix with the $(k,l)$-th
element
$[{{\bf{F}}_{d,ev(od)}}]_{k,l}=e^{-j2{\pi}{\phi}_k{\bar{\tau}}_l/N}$,
where ${\phi}_k \in {\mathcal{I}}_{d,ev(od)}$ and ${\bar{\tau}}_l
\in {\mathcal{I}}_{\hat{\tau}}$, respectively; similarly,
${{\bf{F}}_{p,ev(od)}}$ is a matrix with the $(k,l)$-th element
${{\bf{F}}_{p,ev(od)}}_{k,l}=e^{-j2{\pi}{\theta}_k{\bar{\tau}}_l/N}$,
where ${\theta}_k \in {\mathcal{I}}_{p,ev(od)}$ and ${\bar{\tau}}_l
\in {\mathcal{I}}_{\hat{\tau}}$, respectively;
${\mathcal{I}}_{d,ev(od)}$ denotes the indexes of data tones
allocated to the given user in the even (odd) OFDMA symbols.

Finally, the channel estimation on the data tones is
\begin{equation}
{{\bf{\hat{H}}}_{d,ev(od)}={\bf{G}}_{ev(od)}{\bf{\hat{H}}}_{p,ev(od)}}
\end{equation}
where ${\bf{\hat{H}}}_{p,ev(od)}$ denotes the LS channel estimation
on the pilot tones of even (odd) OFDMA symbols.

\section{Simulation Results}
\label{sec:simu}

The performance of the proposed channel estimation algorithm is
evaluated for a WiMAX system \cite{WiMAX16e} with $BW=10$ MHz,
$N=1024$, $N_{used}=840$, $L_{cp}=128$, $f_c=3.5$ GHz. For
simplicity, data tones are QPSK modulated, and no forward error
coding is applied. ITU Vehicular A channels \cite{ITUR1225} is
adopted, which consists of six individually faded taps with relative
delays as [0, 310, 710, 1090, 1730, 2510] ns and average power as
[0, -1, -9, -10, -15, -20] dB. The Jakes's spectrum \cite{Stee92} is
applied to generate the Rayleigh fading channel. Besides, ideal
synchronization is assumed, thus the delay of the first tap is
always zero ($\tau_0=0$).

\begin{figure}[!t]
\centering
\includegraphics[width=3.4in,height=2.3in]{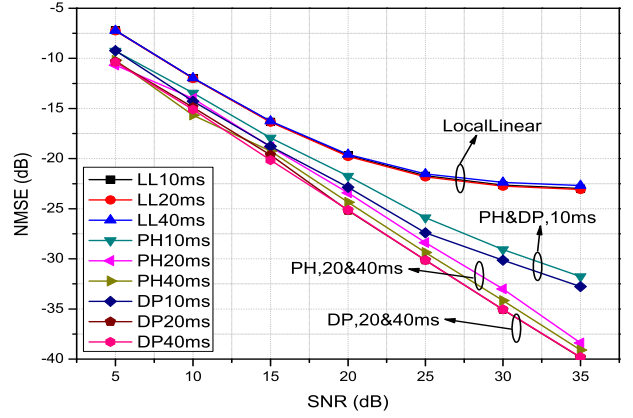}
\caption{NMSE comparison for LL, PH and DP under different $N_t$'s
when $f_d=200$ Hz and $N_{sch}=20$.} \label{fig:fig1}
\end{figure}

The performance of the proposed pilot hopping based channel
estimation algorithm (PH) is compared with the local linear
interpolation algorithm (LL) when uplink virtual MIMO is active. The
algorithm in \cite{Ragh07} which can be called the "doublet-pilots"
algorithm (DP) is also evaluated as a benchmark, and for it, MIMO is
inactive. For LL, the CFR of a certain data tone is simply equal to
the arithmetic average of the two pilot tones within the same tile,
since the irregular distributed pilot pattern prohibits the global
interpolation. Since the second symbol in a tile contains no pilots,
PH is adjusted to accommodate this situation: the second symbol in
each tile is removed and CFR of this symbol is linearly interpolated
from the first and third symbols within the same tile. There are
three main factors influence the performance of PH and DP
dominantly: the number of observed OFDMA symbols ($N_t$), Doppler
($f_d$), and the number of allocated subchannels \footnote{Each
subchannel consists of six tiles random distributed in frequency
domain.} ($N_{sch}$).

First, different $N_t$'s are evaluated to examine the convergence of
PH. In Fig.1, three values of $N_t$, i.e., 10ms, 20ms and 40ms,
which are 96, 192 and 387 OFDMA symbols, correspondingly, are
plotted when $f_d=200$ Hz and $N_{sch}=20$. Obviously, the
performance of LL is independent of $N_t$, and it levels off at high
SNR regime. For PH and DP, no significant error floor can be
observed. When $N_t=10$ ms, PH at least has a 2.5dB gain over LL at
low SNR regime, and far better than LL at high SNR regime. Moreover,
PH is about 2dB worse than DP for all $N_t$'s. However, it is worth
noting that for the same $N_t$, the number of pilots available for
DP is two times for PH, since MIMO is active for PH but not for DP.

Fig.2 plots the performances of LL, PH and DP under different
Doppler's, when $N_t=192$ and $N_{sch}=20$. For LL, the performance
gets worse along the increasing of $f_d$, and levels off at high SNR
regime for all $f_d$'s. However, for PH and DP, no significant error
floor can be observed for lower $f_d$'s. Although PH also levels off
when $f_d$ is high, it still outperforms LL over 10dB at high SNR
regime. Moreover, when $f_d$ is lower, the performance difference
between PH and DP is subtle, and when $f_d$ is higher, the NMSE
difference is about 5dB at high SNR regime.
\begin{figure}[!t]
\centering
\includegraphics[width=3.4in, height=2.3in]{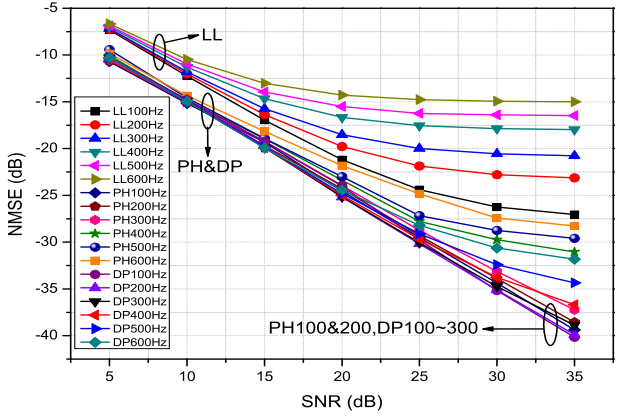}
\caption{NMSE comparison for LL, PH and DP under different $f_d$'s
when $N_t=192$ and $N_{sch}=20$.} \label{fig:fig2}
\end{figure}

Finally, different $N_{sch}$'s are evaluated in Fig.3. From the
figure, it is obvious that the performance difference between PH and
DP is subtle when $N_{sch}\ge20$, and both of them are far better
than LL. When $N_{sch}=10$, PH outperforms LL about 2dB at low SNR
regime and over 8dB at high SNR regime, meanwhile, it is worse than
DP for about 1dB at low SNR regime and 6dB at high SNR regime.

From simulations, we find that PH outperforms LL over a wide range
of SNR's and Doppler's when the number of observed OFDMA symbols and
the number of allocated subchannels are not too small, i.e.,
$N_t\ge96$ and $N_{sch}\ge10$.

\section{Conclusion}
\label{sec:conl} In this paper, we propose an ESPRIT-based channel
estimation algorithm applicable to uplink OFDMA by exploiting pilot
hopping. Through a very simple pilot hopping scheme, the
shift-invariance property based on which ESPRIT is capable is
acquired. Since this special property is derived from a pair of
contiguous OFDMA symbols, no special pilot pattern, e.g., the
doublet pilots in \cite{Ragh07}, is indispensable within one OFDMA
symbol. Hence, the proposed algorithm increases the spectrum
efficiency and eases the design of the pilot pattern. Estimating the
auto-correlation matrix over numbers of OFDMA symbols, the proposed
algorithm outperforms the linear local interpolator within a wide
range of SNR's and Doppler's. Besides, low rank adaptive filter
\cite{Stro96} can be integrated to abate the estimation latency.

\begin{figure}[!t]
\centering
\includegraphics[width=3.4in, height=2.3in]{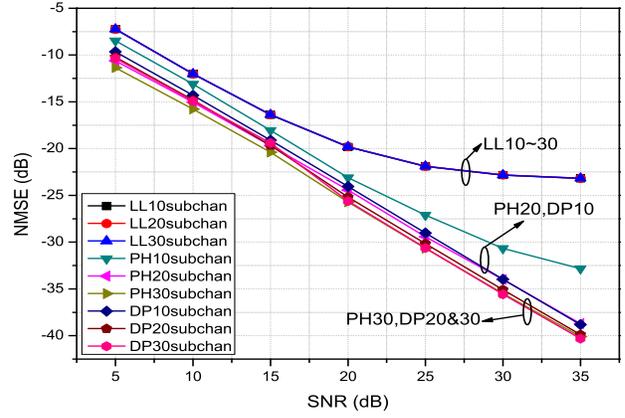}
\caption{NMSE comparison for LL, PH and DP under different
$N_{sch}$'s when $N_t=192$ and $f_d=200$ Hz.} \label{fig:fig3}
\end{figure}

\bibliographystyle{IEEEtran}
\bibliography{IEEEabrv,strings}

\end{document}